\documentclass[12pt, leqno]{article} 
\usepackage[utf8]{inputenc}

\usepackage{geometry}
\usepackage{setspace}
\usepackage{adjustbox}
\usepackage{amsmath}
\usepackage{longtable}
\usepackage{pdfpages}
\usepackage{hhline}
\makeatletter
\renewcommand*\env@matrix[1][*\c@MaxMatrixCols c]{%
  \hskip -\arraycolsep
  \let\@ifnextchar\new@ifnextchar
  \array{#1}}
\makeatother

\usepackage{titlesec} 
 \titleformat{\section}[hang]
  {\large\bfseries\sffamily}
  {\thesection}
  {1em}
  {\MakeUppercase}
\titleformat{\subsection}[hang]
  {\fontsize{15}{15}\bfseries\rmfamily}
  {\thesubsection}
  {1em}
  {}
\titleformat{\subsubsection}[hang]
  {\normalsize\bfseries\sffamily\itshape}
  {\thesubsubsection}
  {1em}
  {}
 
\usepackage{upgreek}
\usepackage{amsthm}
    \theoremstyle{definition}
    
     \theoremstyle{remark}
    
\usepackage{fancyhdr}
\usepackage{float}
\usepackage{placeins}
\usepackage{amssymb}
\usepackage{caption}
\usepackage{commath}
\usepackage{graphicx}
\usepackage{wrapfig}
\usepackage[inline, shortlabels]{enumitem} 
\usepackage{multirow}
\usepackage{pdfpages}  
\usepackage{indentfirst}

\usepackage[english]{babel}
\usepackage{csquotes}
\usepackage{array}
\usepackage{booktabs,caption}
\usepackage[flushleft]{threeparttable}
\usepackage[toc,page]{appendix}
\usepackage[colorlinks=true, allcolors=blue]{hyperref}
\usepackage{pdflscape}
\usepackage{rotating}

\usepackage{tikz}
\usepackage{annotate-equations}

\usepackage{fancyhdr}
\usepackage[authordate,backend=biber,bibencoding=inputenc,autolang=other,url=false,doi=false,eprint=false,numbermonth=false,isbn=false,noibid]{biblatex-chicago} 

\title{How much does SNAP Matter? SNAP’s Effects on Food Security}
\author{Seungmin Lee} 
\date{September 2024}

\begin{document}

\maketitle
\begin{abstract}
    \footnotesize \noindent Supplemental Nutrition Assistance Program (SNAP) aims to improve food security of low-income households in the U.S. A new, continuous food security measure called the Probability of Food Security (PFS), which proxies for the official food security measure but is implementable on longer periods, enables the study of SNAP's effects on the intensive margin. Using variations in state-level SNAP administrative policies as an instrument for individual SNAP participation,  I find that SNAP does not have significant effects on estimated food security on average, both on the entire population and low-income population whom I defined as income is below 130\% of poverty line at least once during the study period. I find SNAP has stronger positive effects on those whose estimated food security status is in the middle of the distribution, but has no significant effects in the tails of the distribution. 
\end{abstract}

\section{Introduction}  \label{intro}

  Food security is defined as access by all people at all times to enough food for an active, healthy life \parencite{world_food_summit_rome_1996}. Food security is a fundamental human right and is associated with a range of well-being outcomes, including child nutrition, mental health and cognitive problems \parencite{gundersen_food_2015}. 12.8\% of households in the U.S. were food insecure in December 2022, and more than one out of ten households have been food insecure every year since 1995, the year the United States Department of Agriculture (USDA) first estimated food security \parencite{rabbitt_household_2023}. More surprisingly, food insecurity is often recurrent (chronic) rather transitory; among households that were food insecure at any point in 2022, 25\% of them were food insecure in almost every month, and among households with ``very low food security'' (the worst food insecurity status), 75\% suffered that status in three or more months, and 25\% of them in almost every month. \parencite{rabbitt_household_2023}

    The Supplemental Nutrition Assistance Program (SNAP), formerly known as the Food Stamp Program, is a federal safety net program designed to reduce poverty and food insecurity among the low income population. SNAP provides benefits to purchase healthy foods at participating food retail outlets. SNAP eligibility and benefit amount are mainly determined by household income. One out of eight people in the U.S. (approximately 41 million) received SNAP benefits in 2022, \$230 per month on average \parencite{usda_supplemental_2023}. Many low-income households' food spending relies heavily on SNAP benefits, implying that the loss of eligibility or a decrease in benefit could have a negative consequence on food security as well as other well-beings, both in the short-term and the long-term. For instance, the gradual state-level phase-out of the SNAP emergency allotment that provided additional benefits in response to the COVID-19 pandemic and ended in March 2023 in the last participating states, is widely perceived to have put SNAP participating households at greater risk of food insecurity, financial insecurity and housing instability \parencite{propel_april_2023}. Monthly surveys of a random sample of SNAP households suggest that the share of households that skipped meals in April 2023 increased by 42\% (to nearly 50\%) in a month, and over 30\% relied at least partially on food pantries for food consumption, the highest ratio since January 2021 \parencite{propel_april_2023}.
    
    SNAP has been politically controversial since it first became a permanent safety net program in 1964 \parencite{bosso_why_2023}. From one perspective, SNAP is essential to protect low-income residents from hunger and poverty, while from another perspective SNAP discourages work among the able-bodied by providing income. These conflicting perspectives have caused SNAP program rules - eligibility and benefits - to undergo considerable changes over the decades, both at state and national levels. For instance, the Personal Responsibility and Work Opportunity Reconciliation Act (PRWORA) of 1996 eliminated SNAP eligibility from most legal immigrants (later restored in 2002), and imposed work requirements and a three-month maximum SNAP benefit periods limit on the able-bodied adults without dependents (ABAWDs), a person aged 18 through 49 who does not have a child under age 18 in their SNAP household and who is fit for work. The 2023 debt ceiling deal as included people aged 50 in the ABAWDs, and will gradually include those aged 51-54 in the next couple of years.

    Research exhibits mixed findings on the effects of SNAP on food security, from positive effects on reducing food insecurity or negative effects of the loss of eligibility \parencite{borjas_food_2004,yen_food_2008,mykerezi_impact_2010,ratcliffe_how_2011,shaefer_supplemental_2013} to null effects \parencite{gundersen_food_2001,gibson-davis_cautionary_2006,chojnacki_randomized_2021}.\footnote{\textcite{schanzenbach_understanding_2023} summarizes the broader SNAP literature, including the effects on other well-being indicators.} However, existing studies focus mostly on the extensive margin (i.e., whether households are food secure or not) rather than on the intensive margin (i.e., how severe household food security is). Only existing studies of SNAP's intensive marginal effects found that SNAP decreases food insecurity by 7\% \parencite{yen_food_2008} and 30-40\% \parencite{mykerezi_impact_2010}.
    
    This limited number of the studies of SNAP's intensive marginal effects is due to the nature of the existing food security measure (Food Security Scale Score, FSSS). FSSS, an official food security measure designed by the USDA. FSSS is a discrete, ordinal measure categorizing food security status as ``food secure'', ``marginally food secure'', ``low food secure (or food insecure)'' or ``very low food secure (or very food insecure)'', depending on the number of questions respondents affirmatively answered to the Household Food Security Survey Module (HFSSM). The USDA estimates the official food insecurity prevalence rate based on the HFSSM administered in the Current Population Survey (CPS) every December. FSSS's discrete nature has limited researchers from studying the SNAP's intensive marginal effects on food insecurity. However, it is important to know whether SNAP reduce the level and severity of food insecurity, on those who remain food insecure. 

    In this paper, I investigate the effects of the SNAP on food security over 17 years, using longitudinal individual-level data from the Panel Study of Income Dynamics (PSID) over 9 rounds from 1997 to 2015. I assess household-level food security using the Probability of Food Security (PFS), a food insecurity measure defined as the estimated probability that a household’s predicted food expenditures equal or exceed the minimum cost of a healthful diet, reflected in the USDA's Thrifty Food Plan (TFP)  which anchors SNAP benefits (\cite{lee_food_2023}, LBH hereafter). LBH established that the PFS serves as a good proxy for the USDA's official food security measure, Food Security Scale Score (FSSS), but unlike the FSSS, the PFS can be implemented in longer panel data sets, like Panel Study of Income Dynamics (PSID), that have food expenditure and household demographic and socioeconomic data. Furthermore, the PFS is a continuous measure which can be used to estimate SNAP's effects on not only food insecurity incidence (i.e, the extensive margin) but also on food insecurity severity (i.e, the intensive margin) which have not been done in the literature. Furthermore, the PFS can be constructed from the existing panel data to observe household- or individual-level food insecurity over a larger period that has been feasible to date.

    I use variation in state-level SNAP administrative policies for causal identification, as others have successfully done \parencite{yen_food_2008,meyerhoefer_does_2008,ratcliffe_how_2011,kreider_identifying_2012,gregory_does_2015,swann_household_2017,heflin_does_2024}. Legislative changes since 1990, including the 1996 welfare reform and the 2002 Farm Bill, empowered states to implement their own SNAP administrative rules determining eligibility, enrollment and re-enrollment process, such as exempting vehicles from eligibility test and requiring fingerprints from applicants. States have adopted different rules at different times, generating considerable state-level variations over the years. I use USDA's SNAP Policy Index \parencite{stacy_using_2018}, which assesses the generosity of SNAP policies, as an instrument to control for endogenous individual SNAP participation. This identification strategy is based on the hypotheses that SNAP administrative policies are strongly relevant to SNAP participation, and that they affect estimated food security only through SNAP participation. 

    I find that SNAP does not have significant effects on estimated food security on average, both on the entire population and low-income population whom I defined as income is below 130\% of poverty line at least once during the study period. I find SNAP has stronger positive effects on those whose estimated food security status is in the middle of the distribution, but has no significant effects in the tails of the distribution. 

\section{Data}
\subsection{Panel Study of Income Dynamics (PSID)} \label{sec:PSID}
PSID is a nationally representative panel survey of U.S. families. Starting with 18,000 individuals from 4,800 households in 1968, the PSID has surveyed 82,000 individuals from about 9,000 households over 42 waves as of 2021, annually until 1997 and biennially since then. Since its initial survey in 1968, the PSID has followed those surveyed in 1968 as well as those who are genealogically related to them (i.e. children and grandchildren). The PSID sample has remained nationally representative by regularly adjusting survey weights to capture attrition and new immigration, as validated by using various economics indicators, against similar estimates from other nationally representative surveys (\cite{andreski_estimates_2014,li_new_2010,gouskova_comparing_2010,tiehen_food_2020}). The PSID collects the individual-level information (e.g., household role, demographics, socioeconomic status) as well as household-level information (e.g., expenditures, SNAP participation).\footnote{Strictly speaking, the PSID collects information on a ``family'', which differs from ``household'' in the survey. Household is a location-based definition which can include more than one family residing in a single housing unit. However, as of the latest PSID survey wave in 2021, more than 92\% of households consist of a single family. Therefore, I use the term ``household'' synonymously with ``family,'' as is common in the literature.}

 I construct \underline{individual-level} panel data of 83,267 observations from 11,933 individuals over 9 waves (1997-2013). Although food-related outcomes are household-level, we use the individual-level data due to the nature of the PSID data and to ensure consistency with the way the outcome measure is constructed. The PSID assigns a unique ID per individual, but does not assign a unique ID per household over time. If a person has lived in the same household over time, the PSID assigns different household IDs to that person’s household in every survey period, even if there has been no change in the household at all. Second, the PFS is a function of conditioning variables, period and panel data construction methods, implying that different construction methods could yield different PFS estimates. Instead of constructing household-level PFS only for this study, I use the PFS constructed from the individual-level panel data of 40-year period (1979-2019) introduced in \textcite{lee_probability_2024}. I do not include Hawaii and Alaska, which do not have the PFS measure due to the absence of monthly TFP cost, and do not include New Hampshire and a subset of New Jersey, South Dakota, Maine and Rhode Island due to the absence of the Cost of Living Index (COLI) \parencite{council_for_community_and_economic_research_historical_2023} which I use to adjust for spatial variation in the food prices.\footnote{The following state-years are excluded due to the absence of the COLI data: Maine (2007, 2009, 2011, 2013), New Jersey (all but 1999), Rhode Island (2005 to 2013), South Dakota (all but 2009)} I use the PSID's longitudinal individual survey weights for weighted estimates, as the unit of analyses is individual-level. I use weighted estimates as my primary results, and replicate key results without weights in Section \ref{sec:wgt_vs_uwgt}.

 Table \ref{Tab:Tab_1} presents weighted summary statistics of the sample.\footnote{Table \ref{Tab:Tab_1} in the Appendix provides unweighted summary statistics.} The left panel is the entire study sample, the right panel is the subsample whose household income was below 130\% of the federal poverty line (FPL), the income threshold for SNAP eligibility, at least once over the study period. It is important to note that the definition of low-income population is individual-level; among the individuals categorized as low-income, 65\% of the observations had income over 130\% of the FPL, and 77\% of the observations had income over 100\% of the FPL at specific year. 23\% of household units have a female reference person (RP, the official term that replaced ``household head'' in the PSID since 2017), 81\% of RPs are White and 66\% are married. 7\% used SNAP with an average monthly benefit of \$330. The share of SNAP benefit amount on income is 15\% at median, and 63\% at 90th percentile. These shares imply that getting SNAP benefits would increase income by 18\% at median, and by 170\% at 90th percentile. 78\% of the sample is likely to spend at least the TFP cost, and 13\% of the observations have the PFS below a certain cut-off, the threshold I use to define individuals as being food insecure.

\begin{table}[htbp]\centering
\caption{Summary Statistics}
\begin{adjustbox}{width=\textwidth}
\begin{threeparttable}
\begin{tabular}{l*{2}{ccc}}
\hline
     &\multicolumn{3}{c}{(Full sample)}               &\multicolumn{3}{c}{(Low income population)}               \\
                    &        N&          mean&         sd&        N&          mean&         sd\\
\hline
Reference Person	 &       &       &        &    &        &       \\
Female (=1)         &      83,234&        0.23&        0.42&      39,867&        0.38&        0.49\\
Age (years)         &      83,234&       49.30&       16.48&      39,867&       47.01&       17.92\\
White (=1)          &      83,234&        0.81&        0.39&      39,867&        0.69&        0.46\\
Married (=1)        &      83,234&        0.66&        0.47&      39,867&        0.48&        0.50\\
Employed (=1)       &      83,234&        0.71&        0.45&      39,867&        0.61&        0.49\\
Disabled (=1)       &      83,234&        0.19&        0.39&      39,867&        0.25&        0.43\\
Highest educational degree & & & & \\
\hspace{3mm}Less than high school (=1)&      83,234&        0.12&        0.33&      39,867&        0.24&        0.43\\
\hspace{3mm}High school (=1)    &      83,234&        0.35&        0.48&      39,867&        0.42&        0.49\\
\hspace{3mm}College w/o degree (=1)&      83,234&        0.19&        0.39&      39,867&        0.17&        0.37\\
\hspace{3mm}College degree (=1) &      83,234&        0.33&        0.47&      39,867&        0.18&        0.38\\
\hline
Household	 &       &       &        &    &        &       \\
Household size      &      83,234&        2.81&        1.49&      39,867&        2.93&        1.72\\
\% children in household&      83,234&        0.20&        0.25&      39,867&        0.25&        0.27\\
Monthly income per capita (thousands)&      83,234&        3.12&        2.68&      39,867&        1.73&        1.82\\
Food exp (with FS benefit)&      83,234&      315.68&      188.25&      39,867&      264.99&      172.32\\
Received SNAP (=1)  &      83,234&        0.07&        0.25&      39,867&        0.17&        0.38\\
SNAP benefit amount &      10,501&      330.37&      231.69&       9,950&      334.48&      235.22\\
SNAP Policy Index (unweighted)&      83,234&        5.99&        2.02&      39,867&        5.98&        1.99\\
SNAP Policy Index (weighted)&      83,234&        7.39&        1.79&      39,867&        7.37&        1.78\\
PFS                 &      83,234&        0.78&        0.23&      39,867&        0.67&        0.25\\
Food Insecure (=1 if PFS below cut-off)             &      83,234&        0.13&        0.34&      39,867&        0.24&        0.43\\
\hline
\end{tabular}
\begin{tablenotes}
\small
    \item \footnotesize * Including SNAP benefit amount
    \item \footnotesize ** Non-SNAP households are excluded.
    \item \footnotesize Monetary values are converted to Jan 2019 dollars using Jan 2019 Consumer Price Index. Top 1\% values of monetary variables are winsorized. Estimates are weighted using longitudinal individual weight in the PSID.
\end{tablenotes}
\end{threeparttable}
\end{adjustbox}
\label{Tab:Tab_1}
\end{table}

\subsection{SNAP Policy Index} \label{sec:SPI}

SNAP is a federally funded program for which the federal government determines income eligibility and maximum benefit amounts that are uniform across states which administer the program for their residents. The program had little state-level variation initially, but legislative changes since 1990, including the 1996 welfare reform and the 2002 Farm Bill, granted states some autonomy to set their own SNAP administration rules (\cite{stacy_using_2018}). For instance, states can decide to waive certain requirements or make SNAP applications easier to file, each of which could increase SNAP participation, or states can apply stricter eligibility requirements, each of which could discourage SNAP participation by increasing the cost of participation \parencite{currie_explaining_2001}, disproportionately affecting the needier groups whom the program mainly targets (\cite{finkelstein_take-up_2019}). States adopted different administrative rules at different times, and these changes have significantly affected SNAP participation \parencite{ganong_decline_2018,dickert-conlin_downs_2021,heflin_local_2023}. One thing to note is that these state policies do not affect the SNAP benefit amount; the benefit amount is still determined at the federal level. Thus, the effect of state-level SNAP policies affect participation at the extensive margin only.

I use the SNAP Policy Index (SPI), an index capturing the generosity of state administrative rules towards the eligible population developed by \textcite{stacy_using_2018} as a source of exogenous variation in SNAP participation to identify the causal effects of individual-level SNAP participation. The SPI runs 1996 to 2014, constructed from 10 policy variables under four different channels that affect program participation. The first channel is through eligibility; exemption of all (or some) vehicle from the SNAP asset test, `broad-based categorical eligibility (BBCE)', and an eligibility restriction for adult non-citizens. The second channel is through transaction costs; proportion of working households with short re-certification periods (1 to 3 months), simplified reporting, and online application availability. The third channel is through stigma; proportion of state benefits through electronic benefit transfer (EBT) and a fingerprinting requirement. The last channel is through outreach; federally funded ratio or TV advertisement of the program. The index assigns positive (negative) value to the policies that are expected to increase (decrease) the SNAP participation, so a higher index value implies more generous state administrative rules which should be and is positively correlated with the SNAP participation. 

\begin{table}[htbp]\centering
\def\sym#1{\ifmmode^{#1}\else\(^{#1}\)\fi}
\caption{SNAP policy variables and their contributions to the SNAP Policy Index}
\begin{threeparttable}
\begin{tabular}{p{0.5\linewidth}  p{0.15\linewidth}  p{0.15\linewidth}}
\hline\hline
 &Contribution to the Index&Weight   \\
\hline
\underline{Policies affecting eligibility} & &\\
Exempts at least one but not all vehicles from SNAP asset test           &    +     &   1.624    \\
Exempts all vehicles from SNAP asset text          &        +&        1.552\\
Broad-based categorical eligibility (BBCE)         &      +&        1.828\\
Eligibility restrictions for adult non-citizens         &        -&       4.800\\
\underline{Policies affecting transaction costs} & &\\
Proportion of working households with short recertification periods (1-3 months)          &       -&        3.180\\
Simplified reporting          &         +&        1.132\\
Online application availability           &       +&        0.456\\
\underline{Policies affecting stigma} & &\\
Mean proportion of State benefits issued via electronic benefits transfer (EBT)         &         +&        0.276\\
Fingerprinting required during application        &         -&        1.864\\
\underline{Policies affecting outreach} & &\\
Federally funded radio or TV ad        &         +&        0.148\\
\hline\hline
\end{tabular}
\begin{tablenotes}
\small
 \item \footnotesize Source: \textcite{stacy_using_2018}, Table 1
\end{tablenotes}
\end{threeparttable}
\label{Tab:SPI_table}
\end{table}

\textcite{stacy_using_2018} provides SPI in two different versions: unweighted and weighted. The unweighted index is constructed by applying equal weight to all policies in index construction, while the weighted index is constructed by policy-specific weights based upon how much each policy is associated with SNAP participation. Table \ref{Tab:SPI_table} provides the list of state administrative policies, their contribution to the SPI, and the weights used to construct the weighted SPI. Generous (restrictive) policies associated with greater (lesser) SNAP participation are marked as plus (minus) sign in the ``Contribution to the Index'' column. The unweighted index is constructed by summing up the number of generous policies adopted minus the number of restrictive policies adopted. If a state adopts all generous policies but none of the restrictive policies, the unweighted SPI would be six (The first two policies affecting eligibility are mutually exclusive). If a state adopts all restrictive policies but none of the restrictive policies, the unweighted SPI would be -3. Then the final unweighted index is scaled to vary from 1 to 10 by adding 4. The weight is the estimated contribution of each policy on SNAP participation, and is used to construct the weighted SPI which is also scaled to vary from 1 to 10.\footnote{\textcite{stacy_using_2018} provide the full detail of the imputation of weights and the construction of the weighted SPI.} The weighted and unweighted SPIs are very strongly correlated with a Pearson correlation of 0.95 per \textcite{stacy_using_2018}). I use the weighted index as a source of exogenous variation to capture relative importance of each policy, and replicate the key results with unweighted index in the appendix. 

\textcite{stacy_using_2018} provides SPI in two different versions: unweighted and weighted. The unweighted index is constructed by applying equal weight to all policies in index construction, while the weighted index is constructed by policy-specific weights based upon how much each policy is associated with SNAP participation. Table \ref{Tab:SPI_table} provides the list of state administrative policies, their contribution to the SPI, and the weights used to construct the weighted SPI. Generous (restrictive) policies associated with greater (lesser) SNAP participation are marked as plus (minus) sign in the ``Contribution to the Index'' column. The unweighted index is constructed by summing up the number of generous policies adopted minus the number of restrictive policies adopted. If a state adopts all generous policies but none of the restrictive policies, the unweighted SPI would be six (The first two policies affecting eligibility are mutually exclusive). If a state adopts all restrictive policies but none of the restrictive policies, the unweighted SPI would be -3. Then the final unweighted index is scaled to vary from 1 to 10 by adding 4. The weight is the estimated contribution of each policy on SNAP participation, and is used to construct the weighted SPI which is also scaled to vary from 1 to 10.\footnote{\textcite{stacy_using_2018} provide the full detail of the imputation of weights and the construction of the weighted SPI.} The weighted and unweighted SPIs are very strongly correlated with a Pearson correlation of 0.95 per \textcite{stacy_using_2018}). I use the weighted index as a source of exogenous variation to capture relative importance of each policy, and replicate the key results with unweighted index in the appendix. 

\begin{figure}[htbp]
\centering
\begin{minipage}{1.0\textwidth} 
\includegraphics[width=\linewidth]{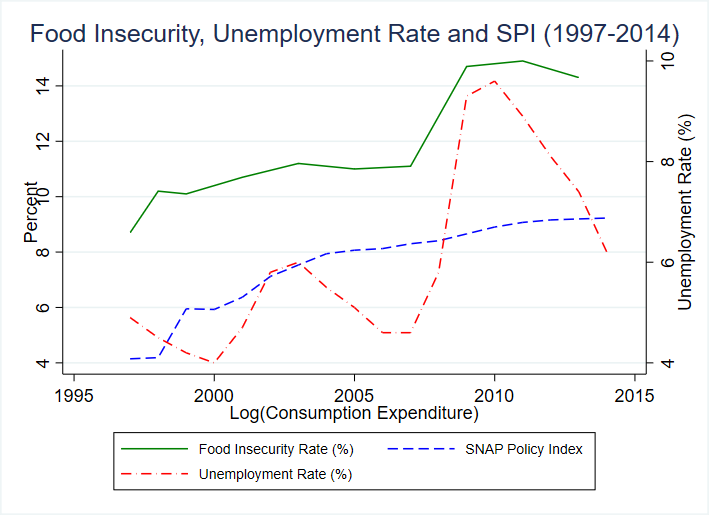}
\end{minipage}
\caption{The SPI and Macroeconomic Indicators, 1996-2014}
\label{fig:SPI_trend}
\end{figure}

\begin{figure}[htbp]
\centering
\begin{minipage}{1.0\textwidth} 
\includegraphics[width=\linewidth]{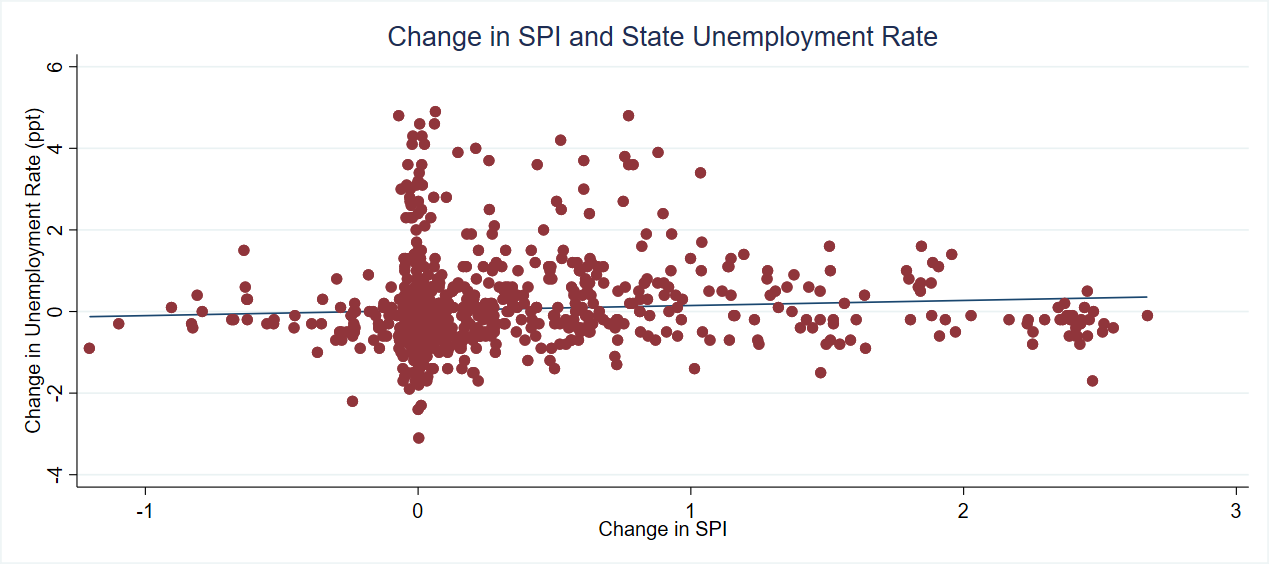}
\end{minipage}
\caption{Change in the SPI and state unemployment rates, 1996-2014}
\label{fig:change_in_SPI_unemp_rate}
\end{figure}

Figure \ref{fig:SPI_trend} shows annual trends of the SPI and two macroeconomic outcomes - official nationwide food insecurity rate and unemployment rate, 1997 to 2014. The SPI was low in 1997, immediately after the 1996 welfare reform which restricted SNAP participation, but gradually increased until 2014. At state-level, the average annual change in SPI is 0.30, and is 0.47 for the years when states relaxed any of the 4 policies affecting eligibility criteria, which accounts for 25\% of the state-year level observations. As of 2014, 14 states (Alabama and 13 others) have the highest SPI (8.8), and Alaska (6.5), Wyoming (6.6) and Indiana (6.8) were the states with the lowest SPI. In terms of within-state change over time, the SPI increased the most in California (3.7 to 8.6) and New York (2.3 to 8.6) \parencite{stacy_using_2018}. These intertemporal variations within states capture greater variations compared to the interstate variations (st.dev 1.63 vs st.dev 1.05). While the U.S. recorded high unemployment and food insecurity during the Great Recession, there was no major change in the SPI trend. Figure \ref{fig:change_in_SPI_unemp_rate} shows that the change in the SPI appears to be uncorrelated with the change in state macroeconomic status reflected in the unemployment rate from 1996 to 2014. The correlation coefficients between those two changes are near zero (0.07) and the null hypothesis of zero effects cannot be rejected at 1\% confidence interval (p-value: 0.04). These findings show that the states did not adjust their administrative policies in response to macroeconomic status, supporting the exogeneity of state SNAP policies necessary for the SPI to provide a defensive instrument for endogenous individual-level SNAP participation.

\section{Empirical Strategy}   \label{emp_strategy}

\subsection{The Probability of Food Security} \label{FSD_measures}

I estimate households' food security status using the PFS measure introduced by LBH. The PFS is the estimated probability that an individual $i$ will have food expenditure greater than or equal to $\underline{W_{it}}$, the TFP cost of the household in which the individual $i$ lives in year $t$, conditional upon the set of covariates $\Theta$. 

I construct the PFS as in \textcite{lee_food_2023}, following three steps introduced in \textcite{cisse_estimating_2018}, with a couple of changes. First, I regress (per capita) monthly food expenditure of individual $i$ in state $s$ in year $t$ on a polynomial of its prior period value - thereby allowing for nonlinear dynamics - as in equation (\ref{eqn:ch3_cond_mean}).

   \begin{equation}
    \label{eqn:ch3_cond_mean}
       W_{ist}=\sum_{\gamma=1}^{2}\pi_{\gamma}W_{is,t-2}^{\gamma}+ \Lambda X_{ist}+ \Omega_{s} + \omega_{t} + \theta_{i} + u_{ist} 
    \end{equation}

\noindent where $X_{ist}$ contains household-level characteristics and state-, year- and individual-fixed effects. To comply with the biennial structure of the PSID since 1997, I include the food expenditure of two years ago (not of the previous year) as the lagged status. The predicted value of $W_{ihst}$,$\hat{W}_{ihst}$, is the estimated conditional mean of $W_{ist}$.

Once the conditional mean of food expenditure is estimated from the equation (\ref{eqn:ch3_cond_mean}), the second step is estimating the conditional variance of $W_{ist}, V[W_{ist}]$. Given a mean zero error term \(E[u_{ist}]=0\), we can estimate it by regressing the squared residual from the equation (\ref{eqn:ch3_cond_mean}) on the same set of covariates as equation (\ref{eqn:ch3_cond_var}) below. The absolute value of the predicted $\hat{u}^2$, $|\hat{\sigma}^2|$, is the conditional variance of monthly household food expenditure per capita (\(V[W_{ist}]=E[|\hat{u}^2_{ist}|]=|\hat{\sigma}^2_{ist}\)|)\footnote{Although $\hat{u}^2$ is non-negative, its predicted value $\hat{\sigma}^2$ is not necessarily negative, which is why we use the absolute value.}.

     \begin{equation}
    \label{eqn:ch3_cond_var}
       \hat{u}^2_{ist}=\sum_{\gamma=1}^{2}\Pi_{\gamma}W_{is,t-2}^{\gamma}+ \lambda X_{ist}+ \Delta_{s} + \delta_{t} + \Theta_{i} + \eta_{ist} 
    \end{equation}
    
The third and the last step is to impose an assumption that $W_{ist}$ follows a specific probability distribution and construct the distribution parameters using the method of moments. As in LBH, I assume $W_{ist}$ follows Gamma distribution as a benchmark distribution since it is non-negative. Then I can calibrate Gamma distribution parameters as $\left(\alpha=\frac{\hat{W}_{it}^{2}}{|\hat{\sigma}_{it}^{2}|},\beta=\frac{|\hat{\sigma}_{it}^{2}|}{\hat{W}_{it}}\right)$. Then the PFS is defined as one minus the conditional cumulative distribution function (CDF) in equation (\ref{eqn:ch3_PFS_function})

    \begin{equation}
    \label{eqn:ch3_PFS_function}
       PFS_{it} = Pr(W_{it}\geq\underline{W_{it}}|\Theta) = 1 - F_{W_{it}}(\underline{W_{t}}|\Theta)\in\left[0,1\right]
    \end{equation}

There are three differences in constructing the PFS between LBH and this paper. First, while the LBH did not adjust for spatial variation in food prices, I adjusted the TFP cost $\underline{W_{it}}$ to account for spatial variation in food prices. The TFP cost does not consider spatial variation in food prices that is strongly associated with regional variations in food security and SNAP purchasing power \parencite{gregory_high_2013,christensen_local_2020,davis_are_2020}. The PFS could under- or over-estimate food security depending on relative food prices without spatial food price variation adjustments. I adjusted the TFP cost based on the Cost of Living Index (COLI) developed by the Council for Community and Economic Research \parencite{council_for_community_and_economic_research_historical_2023}. COLI is a quarterly, metropolitan statistical area (MSA)-level index capturing the relative prices in different categories such as groceries and housing. COLI is constructed in a way that the U.S. national average index equals 100, and the higher the index, the higher the relative prices. I constructed the state-year-level COLI (grocery) index by imputing the state-year-level average. COLI varies from 88 to 166 over the study period. I adjusted the TFP cost by multiplying the TFP cost by COLI divided by 100. Second, while the LBH used a generalized linear model (GLM) logit link regression under Gamma distributional assumption in equation (\ref{eqn:ch3_cond_mean}) and (\ref{eqn:ch3_cond_var}), I use Poisson quasi-MLE which is consistent for any non-negative response variables \parencite{wooldridge_distribution-free_1999}. Third, the LBH included state and year fixed effects, I include state, year and individual fixed effects.

\begin{figure}[htbp]
\centering
\begin{minipage}{0.8\textwidth} 
\includegraphics[width=\linewidth]{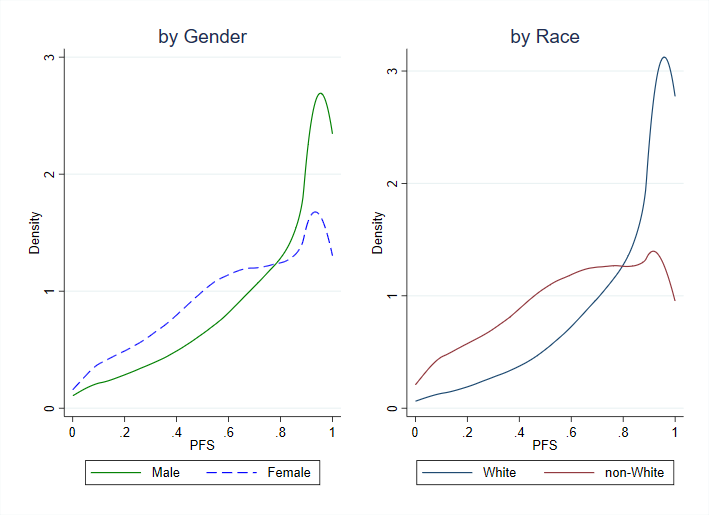}
\end{minipage}
\caption{Kernel Density Plots of the PFS}
\label{fig:PFS_kdensity_bygroup}
\end{figure}

Table \ref{Tab:Ch3_PFS_association} shows the associations between estimated PFS and household-level characteristics. The average PFS in the sample is 0.78 (full sample) / 0.67 (low-income population). The PFS is associated positively with education, employment, income, and negatively with RP being female, having a physical disability, household size. Figure \ref{fig:PFS_kdensity_bygroup} shows kernel density plots of the PFS by different subgroups, where vulnerable groups (women and non-White) are more concentrated in lower PFS. These relationships are intuitive and consistent with the prior literature. The negative associations between SNAP status and the PFS across all specifications imply self-selection into SNAP participation. Furthermore, from the coefficients on monthly income per capita in Table \ref{Tab:Ch3_PFS_association}, I computed semi-elasticity of income on PFS from column (2) and (4), and found 1\% of increase in per capita income increases PFS by 0.027 (full sample) / 0.035 (low-income population). Considering SNAP benefit increases income by 18\% at median and but 170\% at 90th percentile, SNAP benefit would increase PFS by 0.5 at median and 4.6 at 90th on full sample, and by 0.6 at median and 6.0 at 90th on low-income population. These estimated marginal effect size of SNAP benefit on PFS could possibly imply that the causal effect of SNAP on PFS would be very small.

\begin{table}[htbp]\centering
\def\sym#1{\ifmmode^{#1}\else\(^{#1}\)\fi}
\caption{PFS and Household Characteristics}
\begin{adjustbox}{width=\textwidth}
\begin{threeparttable}
\begin{tabular}{l*{4}{c}}
\hline\hline
    &\multicolumn{2}{c}{Full sample}&\multicolumn{2}{c}{Low income population}\\
        &\multicolumn{1}{c}{(1)}&\multicolumn{1}{c}{(2)}&\multicolumn{1}{c}{(3)}&\multicolumn{1}{c}{(4)}\\
    &\multicolumn{1}{c}{PFS}&\multicolumn{1}{c}{PFS}&\multicolumn{1}{c}{PFS}&\multicolumn{1}{c}{PFS}\\
    
\hline
Individual         &     &      &     &      \\
Female (=1)         &      -0.009\sym{***}&       0.000         &      -0.004         &       0.000         \\
                    &      (0.00)         &         (.)         &      (0.00)         &         (.)         \\
Age (years)         &      -0.000         &      -0.003\sym{**} &      -0.001\sym{***}&      -0.006\sym{**} \\
                    &      (0.00)         &      (0.00)         &      (0.00)         &      (0.00)         \\
College degree (=1) &       0.002         &       0.002         &      -0.004         &      -0.010\sym{**} \\
                    &      (0.00)         &      (0.00)         &      (0.01)         &      (0.00)         \\
Reference Person         &     &      &     &      \\
Female (=1)         &      -0.033\sym{***}&      -0.048\sym{***}&      -0.041\sym{***}&      -0.053\sym{***}\\
                    &      (0.00)         &      (0.00)         &      (0.01)         &      (0.00)         \\
Age (years)         &       0.000\sym{***}&       0.001\sym{***}&       0.000\sym{***}&       0.001\sym{***}\\
                    &      (0.00)         &      (0.00)         &      (0.00)         &      (0.00)         \\
White (=1)          &       0.089\sym{***}&       0.006\sym{**} &       0.061\sym{***}&       0.012\sym{***}\\
                    &      (0.00)         &      (0.00)         &      (0.00)         &      (0.00)         \\
Married (=1)        &       0.058\sym{***}&       0.017\sym{***}&       0.030\sym{***}&       0.008\sym{**} \\
                    &      (0.00)         &      (0.00)         &      (0.00)         &      (0.00)         \\
Employed (=1)       &       0.039\sym{***}&       0.039\sym{***}&       0.034\sym{***}&       0.049\sym{***}\\
                    &      (0.00)         &      (0.00)         &      (0.00)         &      (0.00)         \\
Disabled (=1)       &      -0.031\sym{***}&      -0.020\sym{***}&      -0.023\sym{***}&      -0.018\sym{***}\\
                    &      (0.00)         &      (0.00)         &      (0.00)         &      (0.00)         \\
College degree (=1) &       0.036\sym{***}&       0.017\sym{***}&       0.029\sym{***}&       0.017\sym{***}\\
                    &      (0.00)         &      (0.00)         &      (0.00)         &      (0.00)         \\
\hline
Household        &     &      &     &      \\
Household size      &      -0.060\sym{***}&      -0.057\sym{***}&      -0.058\sym{***}&      -0.064\sym{***}\\
                    &      (0.00)         &      (0.00)         &      (0.00)         &      (0.00)         \\
\% children in household&       0.039\sym{***}&       0.009\sym{***}&       0.037\sym{***}&       0.022\sym{***}\\
                    &      (0.01)         &      (0.00)         &      (0.01)         &      (0.01)         \\
Monthly income per capita (thousands)&       0.021\sym{***}&       0.009\sym{***}&       0.032\sym{***}&       0.020\sym{***}\\
                    &      (0.00)         &      (0.00)         &      (0.00)         &      (0.00)         \\
Received SNAP (=1)  &      -0.115\sym{***}&      -0.082\sym{***}&      -0.094\sym{***}&      -0.079\sym{***}\\
                    &      (0.00)         &      (0.00)         &      (0.00)         &      (0.00)         \\
Constant            &       0.749\sym{***}&       0.945\sym{***}&       0.745\sym{***}&       0.972\sym{***}\\
                    &      (0.00)         &      (0.05)         &      (0.01)         &      (0.09)         \\
\hline
N                   &     82,587         &      82,192         &      39,608         &      39,445         \\
R$^2$               &        0.39         &        0.92         &        0.36         &        0.90         \\
Mean PFS            &         0.78         &        0.78         &        0.67         &        0.67         \\
Individual FE       &           N         &           Y         &           N         &           Y         \\
\hline\hline
\end{tabular}
\begin{tablenotes}
\footnotesize
 \item Note: State and Year FE are included. Base category is a male/White/single/male/no college degree/not employed/not disabled.
\end{tablenotes}
\end{threeparttable}
\end{adjustbox}
\label{Tab:Ch3_PFS_association}
\end{table}

\begin{figure}[htbp]
\centering
\begin{minipage}{0.8\textwidth} 
\includegraphics[width=\linewidth]{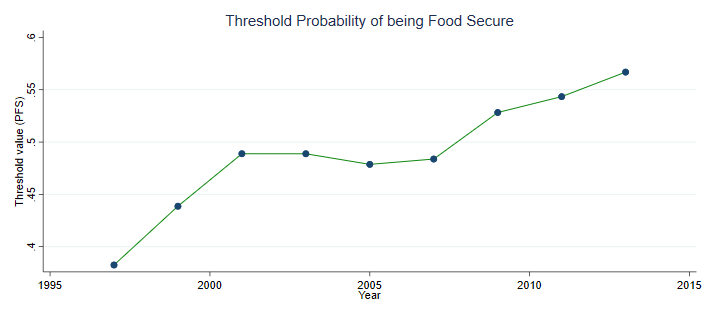}
\end{minipage}
\caption{Threshold Probabilities of being Food Secure, 1997-2013}
\label{fig:PFS_cutoff_9713}
\end{figure}

To determine whether an individual is food secure or not measured by the PFS, I need a threshold probability such that an individual is categorized as food insecure if the PFS is below the threshold. I set year-specific threshold probability in a way that the share of food insecure individuals in the study sample matches the annual individual food insecurity prevalence rate the USDA has reported. Cut-off probabilities vary from 0.38 to 0.57 with the average value of 0.49, as shown in Figure \ref{fig:PFS_cutoff_9713}.




\subsection{Identification Strategy} \label{sec:id_strategy}

I estimate the effects of SNAP participation on food security outcomes, including the PFS and binary food insecurity status (=1 if PFS is below cut-off probability), using a linear two-way fixed effects (TWFE) model. 
\begin{equation}
    \label{eqn:Y_on_SNAP}
    Y_{ist} =\beta_{0} + \beta_{1} SNAP_{ist}+\beta_{2} X_{ist}+ \varphi_{t} + \gamma_{i} + \zeta_{ist}
\end{equation}
\noindent where $Y_{ist}$ is an outcome of interest of an individual $i$ in state \emph{s}, in year \emph{t}, regressed on a vector of covariates $X$, year fixed effect $\varphi_{t}$, and individual-level fixed effect $\gamma_{i}$. The parameter of interest is $\beta_{1}$, the effect of binary SNAP participation status on the estimated food security outcome. Since state SNAP policies do not affect the SNAP benefit amounts, which are federally determined, I do not study the intensive marginal effect of the SNAP benefit amounts on $Y_{ist}$. 

I control for selection into SNAP participation using the 2SLS estimator of $\beta_{1}$, using exogenous variation in SNAP administrative policies as reflected in the SPI as an instrument. As the first stage, I predict SNAP status on SPI and the same set of covariates and fixed effects.
\begin{equation}
    \label{eqn:1st_stage}
    SNAP_{ist}=\alpha_{0} + \alpha_{1} SPI_{ist}+\alpha_{2} X_{ist}+ \phi_{t} + \tau_{i} + \theta_{ist} 
\end{equation}

Then as the second stage I estimate $\beta_{1}$ in equation (\ref{eqn:Y_on_SNAP}) after replacing $SNAP_{ist}$ with the predicted SNAP status $\widehat{SNAP_{ist}}$. To further analyze heterogeneous effects of SNAP on the estimated food security, I add an interaction term of SNAP and a dummy of subgroup as in the equation (\ref{eqn:Y_on_SNAP_hetero}) below.
\begin{equation}
    \label{eqn:Y_on_SNAP_hetero}
    Y_{ist}=\gamma_{0} + \gamma_{1} SNAP_{ist}+ \gamma_{2} SNAP_{ist} \times G_{ist} + \gamma_{3} X_{ist}+ \varphi_{t} + \tau_{i} + \zeta_{ist}
\end{equation}
\begin{equation}
    \label{eqn:1st_stage_hetero}
    SNAP_{ist}=\rho_{0} + \rho_{1} SPI_{ist}+ \rho_{2} SPI_{ist} \times G_{ist} + \rho_{3} X_{ist}+ \varphi_{t} + \tau_{i} + \zeta_{ist}
\end{equation}
where $G_{ist}$ is an indicator dummy of a subgroup and $\gamma_{2}$ captures differential effects. I use three subgroups based on the gender, race and educational attainment of RP. Since $SNAP \times G$ is also endogenous, Equation (\ref{eqn:1st_stage_hetero}) is the associated first-stage regression where $SPI \times G$ is an added instrumental variable to estimate 2SLS of $\gamma_{2}$.

An important note in making inference is that both $Y_{ist}$ and $\widehat{SNAP_{ist}}$ are predicted variables. PFS is a predicted probability relative to an unobserved true probability, and $\widehat{SNAP_{ist}}$ is predicted SNAP participation status of true SNAP status. Measurement error in PFS is caused by measurement errors in conditional mean ($\hat{W}_{ihst}$ in equation \ref{eqn:ch3_cond_mean}) and conditional variance ($\hat{u}^2$ in equation \ref{eqn:ch3_cond_var}), which are caused by error term $u_{ist}$ and $\eta_{ist}$ in equation \ref{eqn:ch3_cond_mean} and \ref{eqn:ch3_cond_var}, respectively. Since those error terms are additive, zero mean value by assumption, and uncorrelated with independent variables by classic strict exogeneity assumption in OLS, measurement error in PFS would make $\hat{\beta_{1}}$ neither biased nor inconsistent, although less precisely estimated. Measurement error in $\widehat{SNAP_{ist}}$, however, would lead $\hat{\beta_{1}}$ to suffering from attenuation bias. Therefore, hypothesizing SNAP effects on PFS would be non-negative, $\hat{\beta_{1}}0$ would be underestimated and less precisely estimated.

\begin{table}[ht]\centering
\def\sym#1{\ifmmode^{#1}\else\(^{#1}\)\fi}
\caption{Weak IV Test}
\begin{adjustbox}{width=\textwidth}
\begin{threeparttable}
\begin{tabular}{l*{6}{c}}
\hline\hline
&\multicolumn{3}{c}{Full sample}&\multicolumn{3}{c}{Low-income population}\\
&\multicolumn{1}{c}{(1)}&\multicolumn{1}{c}{(2)}&\multicolumn{1}{c}{(3)}&\multicolumn{1}{c}{(4)}&\multicolumn{1}{c}{(5)}&\multicolumn{1}{c}{(6)}\\
&\multicolumn{1}{c}{SNAP (=1)} &\multicolumn{1}{c}{SNAP (=1)} &\multicolumn{1}{c}{SNAP (=1)} &\multicolumn{1}{c}{SNAP (=1)} &\multicolumn{1}{c}{SNAP (=1)}&\multicolumn{1}{c}{SNAP (=1)}\\
\hline
SNAP Policy Index &       0.008\sym{***}&       0.008\sym{***}&       0.005\sym{***}&       0.020\sym{***}&       0.022\sym{***}&       0.017\sym{***}\\
                    &      (0.00)         &      (0.00)         &      (0.00)         &      (0.00)         &      (0.00)         &      (0.00)         \\
\hline
N                   &       82850         &       82850         &       82850         &       39710         &       39710         &       39710         \\
Mean SNAP           &        0.07         &        0.07         &        0.07         &        0.17         &        0.17         &        0.17         \\
Controls and Year FE          &           N         &           Y         &           Y         &           N         &           Y         &           Y         \\
Individual FE       &           N         &           N         &           Y         &           N         &           N         &           Y         \\
F-stat(KP)          &      158.49         &       22.59         &       11.08         &      148.24         &       28.71         &       19.27         \\
\hline \hline
\end{tabular}
\begin{tablenotes}
\footnotesize
 \item Note: Controls include RP’s characteristics (gender, age, age squared race, marital status, disability, and college degree). Estimates are adjusted with longitudinal individual survey weight provided in the PSID. Standard errors are clustered at individual-level.
\end{tablenotes}
\end{threeparttable}
\end{adjustbox}
\label{Tab:Weak_IV_SPI}
\end{table}

Table \ref{Tab:Weak_IV_SPI} presents the survey-weighted estimation results from equation (\ref{eqn:1st_stage}). The F-stat is above 10, the rule of thumb value across all specifications, implying that SPI does not suffer from weak IV problem. SNAP participation is positively associated with administrative policies; one-unit increase in the index is associated with 7\% (0.007/0.07) to 11\% (0.008/0.07) increase in SNAP participation on the full sample (column (1) to (3)), and 10\% (0.017/0.17) to 13\% (0.022/0.17) increase in SNAP participation on the low-income population (column (4) to (6)). Considering the average change of 0.47 in state SPI when states relaxed any policies affecting eligibility criteria, SNAP participation was increased by nearly the half of the size reported in Table \ref{Tab:Weak_IV_SPI} during those periods. In terms of an individual policy, adopting BBCE increased SNAP participation by 12.6\% to 19.8\% (effect sizes multiplied by the contribution of BBCE to SPI). If a person with low-income residing in Wyoming relocates to Alabama in 2014, from the state with the second lowest SPI (6.6) to the state with the highest SPI in 2014 (8.8), the probability of that person participating in SNAP was 22\% to 29\% higher (effect size multiplied by the difference in SPI). These strong associations between SNAP participation and the SPI imply that SNAP administrative policies are relevant to SNAP participation, consistent with the literature suggesting positive (negative) associations between generous (restrictive) SNAP state policies and SNAP participation \parencite{yen_food_2008,meyerhoefer_does_2008,ratcliffe_how_2011,gregory_does_2015,swann_household_2017}. These significant results are robust to using unweighted SPI as reported in Table \ref{Tab:Weak_IV_uwSPI} in the appendix.

\begin{figure}[htbp]
\centering
\includegraphics[width=\linewidth]{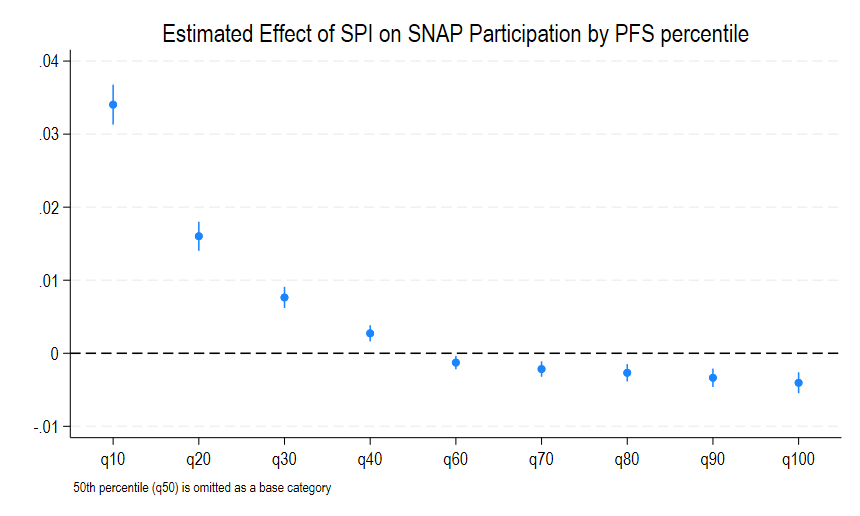}
\caption{Estimated Effect of SPI on SNAP Participation by PFS percentile}
\label{fig:SNAP_on_SPI_qtile}
\end{figure}

Figure \ref{fig:SNAP_on_SPI_qtile} shows the coefficient estimates on G in the equation (\ref{eqn:1st_stage_hetero}) on the full sample, where G are indicators for each PFS quantile. The significant and positive effects on lower quantile show that SPI has greater effects on those with lower estimated food security status, while it has no positive effects on those with higher food security status. These estimates imply that relaxing state SNAP rules is an effective way to increase SNAP participation of the neediest who are likely to be eligible even without eligibility expansion. These inframarginal effects of SPI on increasing SNAP enrollment, so-called ``welcome-mat effect'', is consistent with the finding that expanding state-level eligibility increases SNAP enrollment among the lowest-income individuals \parencite{anders_welfare_2022}. One possible reason for no effects on individuals with higher estimated food security status is that they are already very likely to be SNAP ineligible, thus state policies which do not affect general eligibility have no effects on them.

\begin{table}[htbp]\centering
\def\sym#1{\ifmmode^{#1}\else\(^{#1}\)\fi}
\caption{SNAP on SPI - by Gender, Education and Race}
\begin{adjustbox}{max width=\textwidth}
\begin{threeparttable}
\begin{tabular}{l*{6}{c}}
\hline\hline
&\multicolumn{3}{c}{Full sample}&\multicolumn{3}{c}{Low-income population}\\
&\multicolumn{1}{c}{(1)}&\multicolumn{1}{c}{(2)}&\multicolumn{1}{c}{(3)}&\multicolumn{1}{c}{(4)}&\multicolumn{1}{c}{(5)}&\multicolumn{1}{c}{(6)}\\
&\multicolumn{1}{c}{SNAP (=1)}&\multicolumn{1}{c}{SNAP (=1)}&\multicolumn{1}{c}{SNAP (=1)}&\multicolumn{1}{c}{SNAP (=1)}&\multicolumn{1}{c}{SNAP (=1)}&\multicolumn{1}{c}{SNAP (=1)}\\
&        b/se         &        b/se         &        b/se         &        b/se         &        b/se         &        b/se         \\
\hline
SNAP Policy Index &       0.005\sym{***}&       0.004\sym{***}&       0.004\sym{***}&       0.019\sym{***}&       0.014\sym{***}&       0.017\sym{***}\\
                    &      (0.00)         &      (0.00)         &      (0.00)         &      (0.00)         &      (0.00)         &      (0.00)         \\
SPI x Female (RP)   &       0.001         &                     &                     &      -0.005         &                     &                     \\
                    &      (0.00)         &                     &                     &      (0.00)         &                     &                     \\
SPI x No HS diploma (RP)&                     &       0.005\sym{***}&                     &                     &       0.009\sym{***}&                     \\
                    &                     &      (0.00)         &                     &                     &      (0.00)         &                     \\
SPI x Non-White (RP)&                     &                     &       0.004\sym{*}  &                     &                     &      -0.002         \\
                    &                     &                     &      (0.00)         &                     &                     &      (0.00)         \\
\hline
N                   &       82850         &       82850         &       82850         &       39710         &       39710         &       39710         \\
F-stat(KP)          &        4.61         &        4.77         &        3.48         &       10.21         &        8.61         &        8.98         \\
\hline\hline
\end{tabular}
\begin{tablenotes}
\small
\item \footnotesize \sym{*} \(p<0.10\), \sym{**} \(p<0.05\), \sym{***} \(p<0.01\)
 \item \footnotesize Note: Controls (RP's gender, age, age squared race, marital status, disability college degree), year FE and individual FE are included in all specifications. Estimates are adjusted with longitudinal individual survey weight provided in the PSID. Standard errors are clustered at individual-level.
\end{tablenotes}
\end{threeparttable}
\end{adjustbox}
\label{Tab:SNAP_on_SPI_hetero}
\end{table}

Table \ref{Tab:SNAP_on_SPI_hetero} shows the estimation results from equation (\ref{eqn:1st_stage_hetero}). Combined with significant positive effects on base group, these estimates show that relaxing state SNAP rules increases SNAP participation regardless of the gender, race and educational attainment of RP. In particular, SPI has great effects for less-educated RP on both full sample and low-income subsample. These significant results imply that the local average treatment effect (LATE) estimated from the 2nd-stage regression captures the effects on individuals with different gender, race, and educational attainment of RP. However, they suffer from weak IV across all specifications (F-stat < 10), implying that we may not be able to capture heterogeneous effects of SNAP on the estimated food security with SPI as an instrument.

\section{Results} \label{sec:results}

\begin{table}[htbp]\centering
\def\sym#1{\ifmmode^{#1}\else\(^{#1}\)\fi}
\caption{Estimated Food Security on SNAP Participation}
\begin{threeparttable}
\begin{tabular}{l*{4}{c}}
\hline\hline
    &\multicolumn{2}{c}{Full sample}               &\multicolumn{2}{c}{Low-income population}               \\
&         OLS         &          2SLS         &         OLS         &         2SLS        \\
&\multicolumn{1}{c}{(1)}&\multicolumn{1}{c}{(2)}&\multicolumn{1}{c}{(3)}&\multicolumn{1}{c}{(4)}\\
\hline
SNAP (=1)           &      -0.118\sym{***}&       0.297         &      -0.123\sym{***}&       0.010         \\
                    &      (0.00)         &      (0.22)         &      (0.00)         &      (0.11)         \\
\hline
N                   &        82850         &       82850         &       39710         &       39710             \\
Mean PFS            &         0.78         &        0.78         &       0.67          &       0.67          \\
\hline \hline
\multicolumn{5}{c}{} \\
\multicolumn{5}{c}{Panel A: PFS} \\
\multicolumn{5}{c}{} \\
\hline\hline
     &\multicolumn{2}{c}{Full sample}               &\multicolumn{2}{c}{Low-income population}               \\
&         OLS         &          IV         &         OLS         &          IV         \\
&\multicolumn{1}{c}{(1)}&\multicolumn{1}{c}{(2)}&\multicolumn{1}{c}{(3)}&\multicolumn{1}{c}{(4)}\\
\hline
SNAP (=1)           &       0.188\sym{***}&      -0.699         &       0.197\sym{***}&      -0.167         \\
                    &      (0.01)         &      (0.47)         &      (0.01)         &      (0.26)         \\
\hline
N                   &       82850         &       82850         &         39710         &       39710         \\
Mean FI            &        0.13         &        0.13         &        0.24         &        0.24         \\
\hline \hline
\multicolumn{5}{c}{} \\
\multicolumn{5}{c}{Panel B: Food Insecurity (=1 if PFS below cut-off)} \\
\multicolumn{5}{c}{} \\
\end{tabular}
\begin{tablenotes}
\footnotesize
 \item Note: All models include control variables, year fixed effects and Mundlak controls. Controls include RP’s characteristics (gender, age, age squared race, marital status, disability, and college degree). Estimates are adjusted with longitudinal individual survey weight provided in the PSID. Standard errors are clustered at individual-level.
\end{tablenotes}
\end{threeparttable}
\label{Tab:PFS_on_FS}
\end{table}

Table \ref{Tab:PFS_on_FS} shows the second-stage estimates, from equation (\ref{eqn:Y_on_SNAP}) with full specification (control variables, time- and individual- fixed effects) where $SNAP$ is replaced with the predicted $\widehat{SNAP}$ from equation (\ref{eqn:1st_stage}). Panel A shows the estimates where $Y$ is the PFS and panel B shows that where $Y$ is a binary indicator equal to 1 when an individual is estimated to be food insecure (PFS below cut-off).

In Panel A, OLS coefficients are negative on both full sample (column (1)) and low-income population (column (3)), reflecting selection into SNAP which causes estimates to suffer from downward bias and omitted variable biases. Column (2) and (4) show that participating in SNAP increases the PFS by 38\% (0.30/0.78) in the full sample and by 1\% (0.115/0.67) in low-income population. However, I cannot reject the null hypothesis of zero effects in both samples.

In Panel B SNAP participation decreases the likelihood of being food insecure by 70 percentage points in the full sample and by 17\% in the low-income population, but I cannot reject the null hypothesis of zero effects in either sample. These results imply that SNAP does not have positive effects on increasing the estimated food security status of those who are more likely to get SNAP as state SNAP policies became less stringent. The direction and magnitude of these effects are robust to using unweighted SPI as an instrument as reported in Table \ref{Tab:PFS_on_FS_uwSPI} in the appendix.  

\begin{figure}[htbp]
\centering
\begin{minipage}{1.0\textwidth} 
\includegraphics[width=\linewidth]{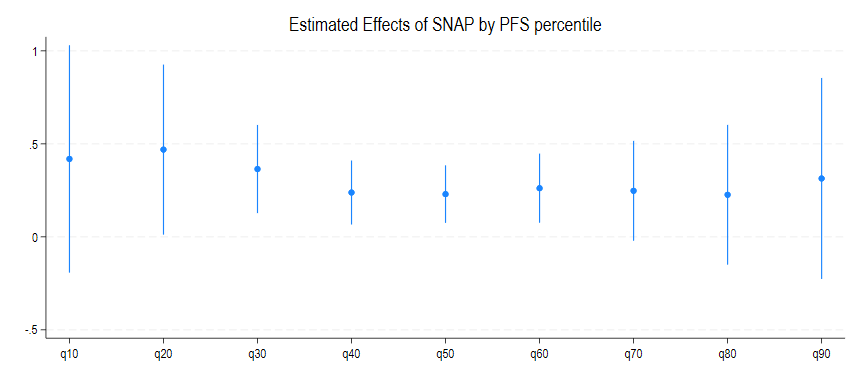}
\end{minipage}
\caption{SNAP Effects on PFS over PFS Distribution}
\label{fig:PFS_qtuile_full}
\end{figure}

To investigate how SNAP effects vary across PFS distribution, I generate quantile estimates of SNAP's effects on PFS.\footnote{Due to the difficulty in running a quantile regression with individual fixed effects using Stata, I instead generated quantile estimates after demeaned dependent and independent variables, which should generate identical estimates by removing individual fixed effects.} Figure \ref{fig:PFS_qtuile_full} shows SNAP effects on the PFS on the full sample over the distribution from 10th percentile (leftmost coefficient) to 90th percentile (rightmost coefficient). All individuals estimated to be food insecure are bottom 20th percentile. This plot implies the following. First, SNAP's effects on increasing food security are stronger on the lower distribution - those with lower food security status - but they are not precisely estimated to be significant. Second, for those extremely food insecure (below 10th percentile), SNAP effect is not as strong as those moderately food insecure, implying that extremely food insecure individuals may suffer from non-income issues (i.e. mental health or homelessness) that cannot be effectively remediable by SNAP benefits. 

\section{Robustness Check} \label{sec:robustness_check}

\subsection{Weighted vs Unweighted Estimates}  \label{sec:wgt_vs_uwgt}

I have reported estimates using the survey weights, which would be more policy relevant as it better represents the full U.S. population. Furthermore, \textcite{solon_what_2015} argued that weighted estimates are preferred over unweighted estimates because (i) they generate more precise estimates by correcting for heteroscedasticity; (ii) they offer consistent estimates after correcting for endogenous sampling; and (iii) they identify average partial effects in the case of heterogeneous effects. However, \textcite{solon_what_2015} also argued that large disparities between unweighted and weighted estimates could imply mis-specification of the model. In addition, if the study sample constructed through the procedure described in Section \ref{sec:PSID} is no longer nationally representative, using survey weights may no longer be appropriate.  \textcite{solon_what_2015} recommended reporting both weighted and unweighted estimates, thus I replicate the main estimates in earlier sections without survey weights. Since the PSID oversamples low-income population, I hypothesize that the (unweighted) estimated effects be greater.

\begin{table}[ht]\centering
\def\sym#1{\ifmmode^{#1}\else\(^{#1}\)\fi}
\caption{Weak IV Test - Unweighted}
\begin{adjustbox}{width=\textwidth}
\begin{threeparttable}
\begin{tabular}{l*{6}{c}}
\hline\hline
&\multicolumn{3}{c}{Full sample}&\multicolumn{3}{c}{Low-income population}\\
&\multicolumn{1}{c}{(1)}&\multicolumn{1}{c}{(2)}&\multicolumn{1}{c}{(3)}&\multicolumn{1}{c}{(4)}&\multicolumn{1}{c}{(5)}&\multicolumn{1}{c}{(6)}\\
&\multicolumn{1}{c}{SNAP (=1)} &\multicolumn{1}{c}{SNAP (=1)} &\multicolumn{1}{c}{SNAP (=1)} &\multicolumn{1}{c}{SNAP (=1)} &\multicolumn{1}{c}{SNAP (=1)}&\multicolumn{1}{c}{SNAP (=1)}\\
\hline
SNAP Policy Index &       0.014\sym{***}&       0.011\sym{***}&       0.007\sym{***}&       0.023\sym{***}&       0.021\sym{***}&       0.017\sym{***}\\
                    &      (0.00)         &      (0.00)         &      (0.00)         &      (0.00)         &      (0.00)         &      (0.00)         \\
\hline
N                   &       82850         &       82850         &       82850         &       39710         &       39710         &       39710         \\
Mean SNAP           &        0.13         &        0.13         &        0.13         &        0.25         &        0.25         &        0.25         \\
Controls and Year FE          &           N         &           Y         &           Y         &           N         &           Y         &           Y         \\
Individual FE       &           N         &           N         &           Y         &           N         &           N         &           Y         \\
F-stat(KP)          &      322.95         &       35.67         &       18.49         &      256.11         &       36.36         &       24.09         \\
\hline \hline
\end{tabular}
\begin{tablenotes}
\footnotesize
 \item Note: Controls include RP’s characteristics (gender, age, age squared race, marital status, disability, and college degree). Standard errors are clustered at individual-level.
\end{tablenotes}
\end{threeparttable}
\end{adjustbox}
\label{Tab:Weak_IV_SPI_uw}
\end{table}

Table \ref{Tab:Weak_IV_SPI_uw} replicates Table \ref{Tab:Weak_IV_SPI} without adjusting survey weights. State policies have greater effects on increasing SNAP participation than in the full sample; from 5\% to 11\% (0.014/0.13). The effects are also larger in the low-income population, but their magnitudes are smaller compared to that in the full sample. These results are consistent with the hypothesis above.

\begin{table}[htbp]\centering
\def\sym#1{\ifmmode^{#1}\else\(^{#1}\)\fi}
\caption{Estimated Food Security on SNAP Participation - Unweighted}
\begin{threeparttable}
\begin{tabular}{l*{4}{c}}
\hline\hline
    &\multicolumn{2}{c}{Full sample}               &\multicolumn{2}{c}{Low-income population}               \\
&         OLS         &          2SLS         &         OLS         &         2SLS        \\
&\multicolumn{1}{c}{(1)}&\multicolumn{1}{c}{(2)}&\multicolumn{1}{c}{(3)}&\multicolumn{1}{c}{(4)}\\
\hline
SNAP (=1)           &      -0.118\sym{***}&      -0.008         &      -0.121\sym{***}&      -0.173\sym{**} \\
                    &      (0.00)         &      (0.12)         &      (0.00)         &      (0.08)         \\
\hline
N                   &        82850         &       82850         &       39710         &       39710             \\
R$^2$               &        0.87         &        0.10         &        0.83         &        0.18         \\
Mean PFS            &        0.72         &        0.72         &        0.62         &        0.62         \\
\hline \hline
\multicolumn{5}{c}{} \\
\multicolumn{5}{c}{Panel A: PFS} \\
\multicolumn{5}{c}{} \\
\hline\hline
     &\multicolumn{2}{c}{Full sample}               &\multicolumn{2}{c}{Low-income population}               \\
&         OLS         &          IV         &         OLS         &          IV         \\
&\multicolumn{1}{c}{(1)}&\multicolumn{1}{c}{(2)}&\multicolumn{1}{c}{(3)}&\multicolumn{1}{c}{(4)}\\
\hline
SNAP (=1)           &       0.192\sym{***}&       0.081         &       0.193\sym{***}&       0.360\sym{*}  \\
                    &      (0.01)         &      (0.26)         &      (0.01)         &      (0.20)         \\
\hline
N                   &       82850         &       82850         &         39710         &       39710         \\
R$^2$               &        0.68         &        0.06         &        0.64         &        0.06         \\
Mean FI            &          0.19         &        0.19         &        0.30         &        0.30         \\
\hline \hline
\multicolumn{5}{c}{} \\
\multicolumn{5}{c}{Panel B: Food Insecurity (=1 if PFS below cut-off)} \\
\multicolumn{5}{c}{} \\
\end{tabular}
\begin{tablenotes}
\footnotesize
 \item Note: All models include control variables, year fixed effects and Mundlak controls. Controls include RP’s characteristics (gender, age, age squared race, marital status, disability, and college degree). Standard errors are clustered at individual-level.
\end{tablenotes}
\end{threeparttable}
\label{Tab:PFS_on_FS_uw}
\end{table}

\begin{figure}[htbp]
\centering
\begin{minipage}{1.0\textwidth} 
\includegraphics[width=\linewidth]{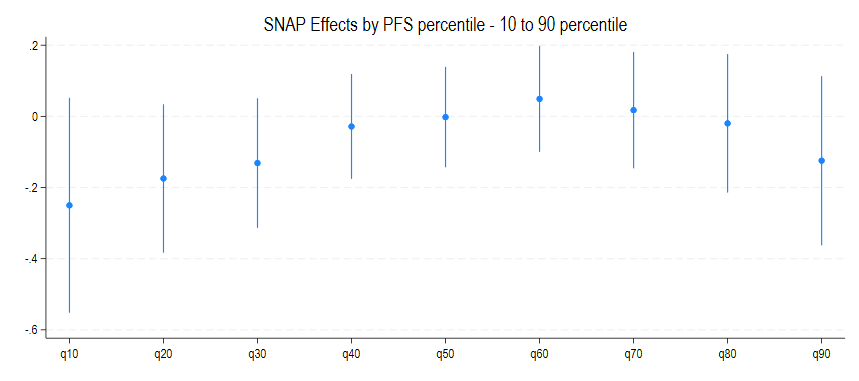}
\end{minipage}
\caption{SNAP Effects on PFS over Distribution - Low-income, unweighted}
\label{fig:PFS_qtile_lnc_uw}
\end{figure}

Table \ref{Tab:PFS_on_FS_uw} replicates Table \ref{Tab:PFS_on_FS} without adjusting survey weights. The results in fact show that SNAP does not have a significant effect in the full sample, consistent with weighted estimates. Surprisingly, SNAP in fact `decreases'' estimated food security (Panel A) and ``increases'' the likelihood of being food insecure on low-income population (Panel B) in these unweighted estimates. Figure \ref{fig:PFS_qtile_lnc_uw} decomposes these counterintuitive effects on low-income population over the distribution of the PFS.\footnote{Figure \ref{fig:PFS_qtile_full_uw} shows unweighted distributional effects on full-sample, whose pattern is robust to the weighted effects described in Figure \ref{fig:PFS_qtuile_full}.} Large negative (but insignificant) effects are concentrated at the bottom of the PFS distribution. If these large discrepancies between weighted and unweighted estimates on low-income population are due to model mis-specificaiton, food security of the low income population is largely determined by something other than the covariates in the model.

\subsection{Non-linear estimation of SNAP Participation}  \label{sec:nonlin_SNAP}

In this section, I replicate \ref{eqn:Y_on_SNAP} using an alternative instrument; non-linearly predicted SNAP participation. Linear estimation of the first-stage equation (\ref{eqn:1st_stage}) resulted in approximately 6\% of the full sample (3\% of the low-income sample) having negative predicted SNAP participation. Although this small share of observations may not significantly bias the estimates, I check the robustness of the results by using an alternative 2SLS estimator using a three-step procedure introduced in \textcite{angrist_mostly_2009} to handle binary endogenous variable. First, I estimate binary SNAP participation on SPI and the set of controls using a logit regression as in equation (\ref{eqn:1st_stage_logit}), and predict SNAP participation status, $\widehat{SNAP_{ist}}$. Second, I estimate binary SNAP participation status on this non-linearly predicted $\widehat{SNAP_{ist}}$ as in equation (\ref{eqn:1st_stage_SNAPhat}) and predict binary SNAP status, $\widehat{\widehat{SNAP_{ist}}}$. Third, I estimate the effects of $\widehat{\widehat{SNAP_{ist}}}$ on the estimated food security outcome using a TWFE model in  equation (\ref{eqn:2nd_stage_SNAPhat}). In other words, the second and the third step are conventional 2SLS estimation using $\widehat{SNAP_{ist}}$ as an instrument.
\begin{equation}
    \label{eqn:1st_stage_logit}
    SNAP_{ist}= f(SPI_{ist}, X_{hst}, \phi_{t}, \tau_{i}) 
\end{equation}
\begin{equation}
    \label{eqn:1st_stage_SNAPhat}
    SNAP_{ist}=\delta_{0} + \delta_{1} \widehat{SNAP_{ist}} +\delta_{2} X_{hst}+ \phi_{t} + \tau_{i} + \theta_{ist} 
\end{equation}
\begin{equation}
    \label{eqn:2nd_stage_SNAPhat}
    Y{ist}=\lambda_{0} + \lambda_{1} \widehat{\widehat{SNAP_{ist}}} +\lambda_{2} X_{hst}+ \phi_{t} + \tau_{i} + \theta_{ist} 
\end{equation}

\begin{figure}[htbp]
\centering
\begin{minipage}{1.0\textwidth} 
\includegraphics[width=\linewidth]{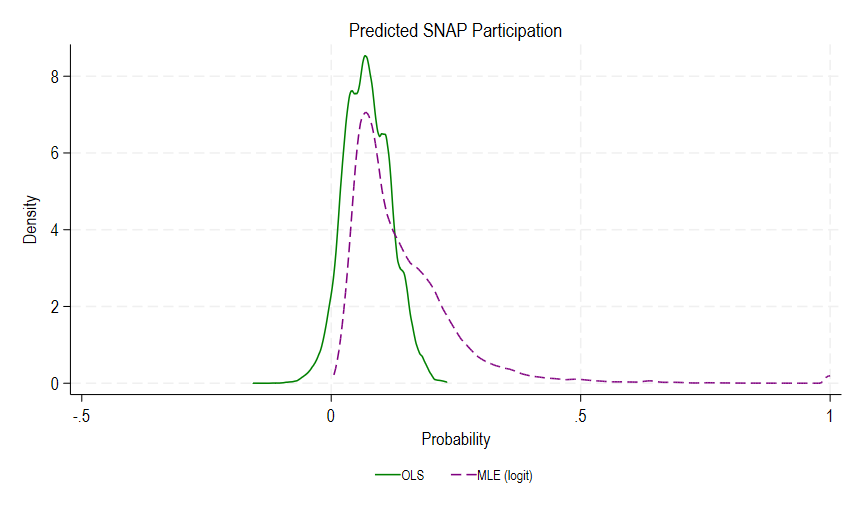}
\end{minipage}
\caption{SNAP Effects on PFS over Distribution}
\label{fig:SNAPhat_OLS_MLE}
\end{figure}

Figure \ref{fig:SNAPhat_OLS_MLE} shows the distribution of the predicted binary SNAP status, linearly estimated in equation (\ref{eqn:1st_stage}) and non-linearly estimated in equation (\ref{eqn:1st_stage_logit}). Compared to the OLS-based predicted SNAP status, logit-based status does not have negative values and is more evenly distributed from value 0 to 1.

\begin{table}[ht]\centering
\def\sym#1{\ifmmode^{#1}\else\(^{#1}\)\fi}
\caption{SNAP Participation on SPI - Logit}
\begin{adjustbox}{max width=\textwidth}
\begin{threeparttable}
\begin{tabular}{l*{6}{c}}
\hline\hline
&\multicolumn{3}{c}{Full sample}&\multicolumn{3}{c}{Low-income population}\\
&\multicolumn{1}{c}{(1)}&\multicolumn{1}{c}{(2)}&\multicolumn{1}{c}{(3)}&\multicolumn{1}{c}{(4)}&\multicolumn{1}{c}{(5)}&\multicolumn{1}{c}{(6)}\\
&\multicolumn{1}{c}{SNAP (=1)} &\multicolumn{1}{c}{SNAP (=1)} &\multicolumn{1}{c}{SNAP (=1)} &\multicolumn{1}{c}{SNAP (=1)} &\multicolumn{1}{c}{SNAP (=1)}&\multicolumn{1}{c}{SNAP (=1)}\\
\hline
SNAP Policy Index &        0.009\sym{***}&       0.008\sym{***}&       0.037\sym{***}&       0.021\sym{***}&       0.024\sym{***}&       0.036\sym{***}\\
        &      (0.00)         &      (0.00)         &      (0.01)         &      (0.00)         &      (0.00)         &      (0.01)         \\
\hline
N                   &       82850         &       82850         &       24218         &       39710         &       39710         &       21465         \\
Mean SNAP           &        0.07         &        0.07         &        0.34         &        0.17         &        0.17         &        0.36         \\
Controls and Year FE          &           N         &           Y         &           Y         &           N         &           Y         &           Y         \\
Individual FE       &           N         &           N         &           Y         &           N         &           N         &           Y         \\
\hline \hline
\end{tabular}
\begin{tablenotes}
\footnotesize
 \item Note: Controls include RP’s characteristics (gender, age, age squared race, marital status, disability, and college degree). Mundlak includes time-average of controls and year fixed effects. Estimates are adjusted with longitudinal individual survey weight provided in the PSID. Standard errors are clustered at individual-level.
\end{tablenotes}
\end{threeparttable}
\end{adjustbox}
\label{Tab:Weak_IV_SNAPhat}
\end{table}

Table \ref{Tab:Weak_IV_SNAPhat} shows the marginal effects of SPI on SNAP participation from the equation (\ref{eqn:1st_stage_logit}), replicating Table \ref{Tab:Weak_IV_SPI}. The estimated marginal effects of SPI on SNAP are very similar in quantity in column (1), (2), (4) and (5). It should be noted that the logit estimation with individual fixed effects cannot be done with the individuals with constant outcome values (SNAP status), dropping significant number of those who never used SNAP (always 0) or always used SNAP (always 1) from the regression sample in column (3) and (6). Since the regression samples are different between Table \ref{Tab:Weak_IV_SPI} and Table \ref{Tab:Weak_IV_SNAPhat} in column (3) and (6), I cannot compare the marginal effects between them to check the robustness.

\begin{table}[htbp]\centering
\def\sym#1{\ifmmode^{#1}\else\(^{#1}\)\fi}
\caption{Estimated Food Security on SNAP Participation - Logit estimation}
\begin{threeparttable}
\begin{tabular}{l*{4}{c}}
\hline\hline
    &\multicolumn{2}{c}{PFS}               &\multicolumn{2}{c}{FI (PFS < cut-off)}               \\
&         Full         &          Low-income         &         Full         &         Low-income        \\
&\multicolumn{1}{c}{(1)}&\multicolumn{1}{c}{(2)}&\multicolumn{1}{c}{(3)}&\multicolumn{1}{c}{(4)}\\
\hline
SNAP (=1)           &       0.664         &      -0.763         &      -4.556         &       7.780         \\
                    &      (3.72)         &      (2.48)         &     (15.63)         &     (31.86)         \\
\hline
N                   &        82850         &       82850         &       39710         &       39710             \\
F-stat (KP)           &         0.1         &        0.1         &       0.1          &      0.1          \\
\hline \hline
\end{tabular}
\begin{tablenotes}
\footnotesize
 \item Note: All models include control variables, year fixed effects and Mundlak controls. Controls include RP’s characteristics (gender, age, age squared race, marital status, disability, and college degree). Estimates are adjusted with longitudinal individual survey weight provided in the PSID. Standard errors are clustered at individual-level.
\end{tablenotes}
\end{threeparttable}
\label{Tab:PFS_on_FS_nonlin}
\end{table}

Table \ref{Tab:PFS_on_FS_nonlin} replicates even columns in Table \ref{Tab:PFS_on_FS} with non-linearly predicted SNAP as an instrument; column (1) and (2) replicate column (2) and (4) in Panel A, and column (3) and (4) replicate those in Panel B. The effects are much greater in magnitude than that reported in Table \ref{Tab:PFS_on_FS} (0.664 vs 0.297 in column 1) but also insignificant. The effects in Column (3) and (4) are too extreme. However, F-stats are extremely weak (0.1) across all specifications, implying that non-linear predicted SNAP status may not be a valid instrument.

Thus, I conclude that the effects of SNAP on the PFS, the main finding of this paper, is robust to the functional form of the first stage.

\section{Conclusion}

This study investigates the effect of SNAP participation on food security over a 17-year period. The study of SNAP's causal effects on food security dynamics at the intensive margin has been limited due to the nature of the official food security measure. I use a new food security measure based on food expenditure and individual- and household demographic and socioeconomic data, which allows me to study SNAP's effects on the level of food security. Using state-level intertemporal variations in SNAP administrative policies as an instrument, I found that relaxing state-level SNAP policies increases SNAP participation, with the strongest effects on those estimated to be very food insecure. However, SNAP does not have significant effects on improving estimated food security status, particularly those with lower estimated food security status. These findings imply that although relaxing SNAP eligibility is an effective way to increase SNAP enrollment of food insecure individuals, SNAP does not significantly improve their estimated food security status.

This study has important limitations, which can be investigated in follow-on research. First, I do not consider the possible misreporting of SNAP participation, which has been increasing \parencite{meyer_household_2015}. Such measurement errors could be both classical due to different recall periods in food expenditure and SNAP status, or non-classical due to stigma. Possible approaches to overcome this limitation would include using SNAP administrative data, partially identifying the effect \parencite{gundersen_partial_2017} or post-startifying survey weights \parencite{jolliffe_food_2023}. Second, I do not investigate the heterogeneous effects of SNAP by previous SNAP redemption pattern.


 

\pagebreak

\begingroup
\setstretch{1.5}
\printbibliography
\endgroup

\clearpage

\begin{appendices}

\setcounter{table}{0}
\setcounter{figure}{0}
\renewcommand{\thefigure}{A\arabic{figure}}
\renewcommand{\thetable}{A\arabic{table}}

\begin{table}[ht]\centering
\def\sym#1{\ifmmode^{#1}\else\(^{#1}\)\fi}
\caption{Weak IV Test - Unweighted SPI}
\begin{adjustbox}{width=\textwidth}
\begin{threeparttable}
\begin{tabular}{l*{6}{c}}
\hline\hline
&\multicolumn{3}{c}{Full sample}&\multicolumn{3}{c}{Low-income population}\\
&\multicolumn{1}{c}{(1)}&\multicolumn{1}{c}{(2)}&\multicolumn{1}{c}{(3)}&\multicolumn{1}{c}{(4)}&\multicolumn{1}{c}{(5)}&\multicolumn{1}{c}{(6)}\\
&\multicolumn{1}{c}{SNAP (=1)} &\multicolumn{1}{c}{SNAP (=1)} &\multicolumn{1}{c}{SNAP (=1)} &\multicolumn{1}{c}{SNAP (=1)} &\multicolumn{1}{c}{SNAP (=1)}&\multicolumn{1}{c}{SNAP (=1)}\\
\hline
SNAP Policy Index (unweighted)&       0.008\sym{***}&       0.007\sym{***}&       0.002         &       0.019\sym{***}&       0.018\sym{***}&       0.009\sym{**} \\
                    &      (0.00)         &      (0.00)         &      (0.00)         &      (0.00)         &      (0.00)         &      (0.00)         \\
\hline
N                   &       82850         &       82850         &       82850         &       39710         &       39710         &       39710         \\
Mean SNAP           &        0.07         &        0.07         &        0.07         &        0.17         &        0.17         &        0.17         \\
Controls and Year FE          &           N         &           Y         &           Y         &           N         &           Y         &           Y         \\
Individual FE       &           N         &           N         &           Y         &           N         &           N         &           Y         \\
F-stat(KP)          &      187.51         &       18.05         &        1.82         &      166.57         &       21.72         &        5.72         \\
\hline \hline
\end{tabular}
\begin{tablenotes}
\footnotesize
 \item Note: Controls include RP’s characteristics (gender, age, age squared race, marital status, disability, and college degree). Estimates are adjusted with longitudinal individual survey weight provided in the PSID. Standard errors are clustered at individual-level.
\end{tablenotes}
\end{threeparttable}
\end{adjustbox}
\label{Tab:Weak_IV_uwSPI}
\end{table}

\begin{table}[htbp]\centering
\def\sym#1{\ifmmode^{#1}\else\(^{#1}\)\fi}
\caption{Estimated Food Security on SNAP Participation - Unweighted SPI as an instrument}
\begin{threeparttable}
\begin{tabular}{l*{4}{c}}
\hline\hline
    &\multicolumn{2}{c}{Full sample}               &\multicolumn{2}{c}{Low-income population}               \\
&         OLS         &          2SLS         &         OLS         &         2SLS        \\
&\multicolumn{1}{c}{(1)}&\multicolumn{1}{c}{(2)}&\multicolumn{1}{c}{(3)}&\multicolumn{1}{c}{(4)}\\
\hline
SNAP (=1)           &      -0.118\sym{***}&       1.192         &      -0.123\sym{***}&       0.301         \\
                    &      (0.00)         &      (1.09)         &      (0.00)         &      (0.27)         \\
\hline
N                   &        82850         &       82850         &       39710         &       39710             \\
R$^2$               &        0.88         &       -6.88         &        0.85         &       -0.88         \\
Mean PFS            &         0.78         &        0.78         &       0.67          &       0.67          \\
\hline \hline
\multicolumn{5}{c}{} \\
\multicolumn{5}{c}{Panel A: PFS} \\
\multicolumn{5}{c}{} \\
\hline\hline
     &\multicolumn{2}{c}{Full sample}               &\multicolumn{2}{c}{Low-income population}               \\
&         OLS         &          IV         &         OLS         &          IV         \\
&\multicolumn{1}{c}{(1)}&\multicolumn{1}{c}{(2)}&\multicolumn{1}{c}{(3)}&\multicolumn{1}{c}{(4)}\\
\hline
SNAP (=1)           &       0.188\sym{***}&      -2.537         &       0.197\sym{***}&      -0.827         \\
                    &      (0.01)         &      (2.30)         &      (0.01)         &      (0.66)         \\
\hline
N                   &       82850         &       82850         &         39710         &       39710         \\
R$^2$               &        0.68         &       -5.86         &        0.66         &       -1.03         \\
Mean FI            &        0.13         &        0.13         &        0.24         &        0.24         \\
\hline \hline
\multicolumn{5}{c}{} \\
\multicolumn{5}{c}{Panel B: Food Insecurity (=1 if PFS below cut-off)} \\
\multicolumn{5}{c}{} \\
\end{tabular}
\begin{tablenotes}
\footnotesize
 \item Note: All models include control variables, year fixed effects and Mundlak controls. Controls include RP’s characteristics (gender, age, age squared race, marital status, disability, and college degree). Estimates are adjusted with longitudinal individual survey weight provided in the PSID. Standard errors are clustered at individual-level.
\end{tablenotes}
\end{threeparttable}
\label{Tab:PFS_on_FS_uwSPI}
\end{table}

 \begin{figure}[t]
    \centering
    \begin{minipage}{0.9\textwidth} 
    \includegraphics[width=\linewidth]{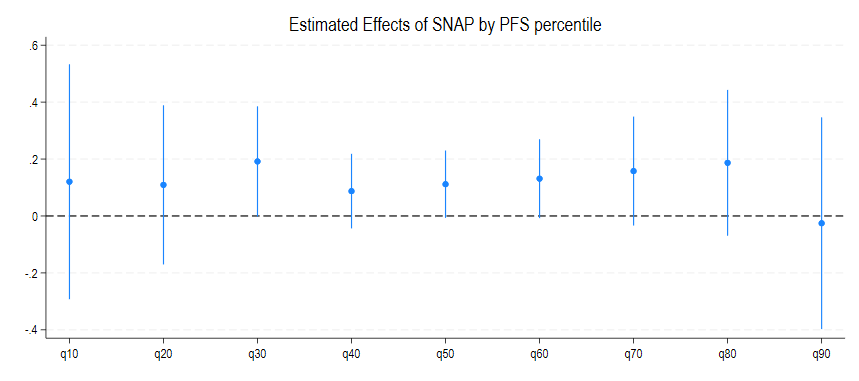}
    \end{minipage}
    \caption{SNAP Effects on PFS over Distribution - Unweighted}
    \label{fig:PFS_qtile_full_uw}
    \end{figure}

\setcounter{table}{0}
\setcounter{figure}{0}
\renewcommand{\thefigure}{B\arabic{figure}}
\renewcommand{\thetable}{B\arabic{table}}

\section{Additional Tables and Figures} \label{additional_tab_fig}


\begin{table}[ht]\centering
\caption{Summary Statistics - unweighted}
\begin{threeparttable}
\begin{tabular}{l*{2}{ccc}}
\hline
     &\multicolumn{3}{c}{(Full sample)}               &\multicolumn{3}{c}{(Low income population)}               \\
                    &        N&          mean&         sd&        N&          mean&         sd\\
\hline
Reference Person	 &       &       &        &    &        &       \\
Female (=1)         &      83,234&        0.30&        0.46&      39,867&        0.45&        0.50\\
Age (years)         &      83,234&       45.83&       15.76&      39,867&       43.57&       16.20\\
White (=1)          &      83,234&        0.59&        0.49&      39,867&        0.41&        0.49\\
Married (=1)        &      83,234&        0.59&        0.49&      39,867&        0.43&        0.49\\
Employed (=1)       &      83,234&        0.72&        0.45&      39,867&        0.62&        0.48\\
Disabled (=1)       &      83,234&        0.17&        0.38&      39,867&        0.21&        0.41\\
Less than high school (=1)&      83,234&        0.16&        0.36&      39,867&        0.26&        0.44\\
High school (=1)    &      83,234&        0.37&        0.48&      39,867&        0.43&        0.49\\
College w/o degree (=1)&      83,234&        0.20&        0.40&      39,867&        0.18&        0.39\\
College degree (=1) &      83,234&        0.27&        0.44&      39,867&        0.14&        0.34\\
\hline
Household size      &      83,234&        3.11&        1.63&      39,867&        3.31&        1.82\\
\% children in household&      83,234&        0.26&        0.27&      39,867&        0.32&        0.28\\
Monthly income per capita (thousands)&      83,234&        2.53&        2.39&      39,867&        1.41&        1.52\\
Monthly food exp per capia&      83,234&      282.02&      178.00&      39,867&      239.22&      160.42\\
Received SNAP (=1)  &      83,234&        0.13&        0.33&      39,867&        0.25&        0.43\\
SNAP benefit amount &      10,501&      365.36&      251.42&       9,950&      369.82&      254.27\\
SNAP Policy Index (unweighted)&      83,234&        5.97&        2.00&      39,867&        6.00&        1.98\\
SNAP Policy Index (weighted)&      83,234&        7.37&        1.81&      39,867&        7.40&        1.80\\
PFS                 &      83,234&        0.72&        0.25&      39,867&        0.62&        0.26\\
FI (=1)             &      83,234&        0.16&        0.37&      39,867&        0.26&        0.44\\
\hline
Outcomes	 &       &       &        &    &        &       \\
PFS                 &      83,234&        0.78&        0.24&      39,867&        0.67&        0.26\\
PFS < 0.5 (=1)&      83,234&        0.15&        0.35&      39,867&        0.26&        0.44\\
\hline
\end{tabular}
\begin{tablenotes}
\small
    \item \footnotesize * Including SNAP benefit amount
    \item \footnotesize ** Non-SNAP households are excluded.
    \item \footnotesize Monetary values are converted to Jan 2019 dollars using Jan 2019 Consumer Price Index. Top 1\% values of monetary variables are winsorized.
\end{tablenotes}
\end{threeparttable}
\label{Tab:Sumstat_uwgt}
\end{table}

\clearpage

\clearpage


\end{appendices}

\end{document}